\documentclass[jkps,twocolumn,fleqn,showpacs,showkeys]{revtex4-1}
\usepackage[pdftex]{graphicx}
\usepackage{amssymb, color}
\usepackage{amsmath}
\usepackage{bm}
\begin{document}
\setcounter{page}{0}
\title[]{Effects of interaction range on the behavior of opinion consensus}
\author{Seungjae \surname{Lee}}
\affiliation{Department of Physics, Chonbuk National University, Jeonju
54896, Korea}
\author{Young Sul \surname{Cho}}
\email{yscho@jbnu.ac.kr}
\author{Hyunsuk \surname{Hong}}
\email{hhong@jbnu.ac.kr}
\affiliation{Department of Physics and Research institute of Physics and Chemistry, Chonbuk National University, Jeonju
54896, Korea}

\today

\begin{abstract}
We have frequently encountered the rapid changes that prevalent opinion of the social community is toppled by a new and opposite opinion against the pre-exiting one. To understand this interesting process, mean-field model with infinite-interaction range has been mostly considered in previous studies 
S. A. Marvel \textit{et al.}, Phys. Rev. Lett. 110, 118702(2012). However, the mean-field interaction range is lack of reality in the sense that any individual cannot interact with all of the others in the community. Based on it, in the present work, we consider a simple model of opinion consensus so-called basic model on the low-dimensional lattices ($d$=1,2) with finite interaction range. The model consists of four types of subpopulations with different opinions: $A, B, AB$, and the zealot of $A$ denoted by $A_c$, following the basic model shown in the work by S. A. Marvel \textit{et al.}. Comparing with their work, we consider the finite range of the interaction, and particularly reconstruct the lattice structure by adding new links when the two individuals have the distance $<\sigma$. We explore how the interaction range $\sigma$ affects the opinion consensus process on the reconstructed lattice structure. We find that the critical fraction of population for $A_c$ 
required for the opinion consensus on $A$ shows different behaviors in the small and large interaction ranges. Especially, the critical fraction for $A_c$ increases with the size of $\sigma$ in the region of small interaction range, which is counter-intuitive: When the interaction range is increased, not only the number of nodes affected by $A_c$ but also that affected by $B$ grows, which is believed to cause the increasing behavior of the critical fraction for $A_c$. We also present the difference of dynamic process to the opinion consensus between the regions of small and large interaction ranges.
\end{abstract}

\pacs{89.65.-s, 89.65.Ef}

\keywords{Opinion dynamics, Social system and Interaction range}

\maketitle

\section{INTRODUCTION}

In the social history, we have encountered various types of social changes induced by conflicts between two opposite opinions. These changes in social systems usually are known to be rapid and radical. Sometimes these changes have been ignited by a small number of people contrary to popular belief, for example, the French Revolution, 
the American civil rights movement~\cite{Morris}, the paradigm shifts in science~\cite{Kuhn, Chen, Bettencourt}, and rapid spreading of political campaigns~\cite{Gerring}.

Many scientists and sociologists have studied these radical changes driven 
by a small subpopulation~\cite{Hong,Xie,Vazquez1,Vazquez2,Centola,Gekle,Galam,Bujalski,Mahmoodi,Chenoweth}. 
For example, Erica Chenoweth revealed in her book~\cite{Chenoweth} that it may only take $3.5\%$ of the population 
to topple a dictator by nonviolent resistances.
To understand these phenomena, various models have been suggested.
In Ref.\cite{Hong}, the authors suggested the basic model. In this model, most of the nodes are initially set to have the same opinion
and the other nodes are zealots of the opposite opinion.  
They studied this model analytically in the mean-field limit and found that a small fraction of subpopulation
of zealots can raise the consensus on their opinion.
Moreover, this consensus emerges drastically (discontinuously) as the number of the zealots exceeds its critical value.
However, people in the real world do not interact with all the others contrary to the mean-field limit.

In this work, we study the basic model on one- and two-dimensional Euclidean spaces, allowing finite-range interactions
to describe the finite-range interactions in real systems. To be specific, we consider one- and two-dimensional Euclidean lattices
and add link between each pair of sites closer than a criterion distance.
On this network structure, we perform numerical simulations, and investigate the effects of the interaction range on the opinion consensus behavior.

\section{Model}

\begin{figure}
\includegraphics[width=1.0\linewidth]{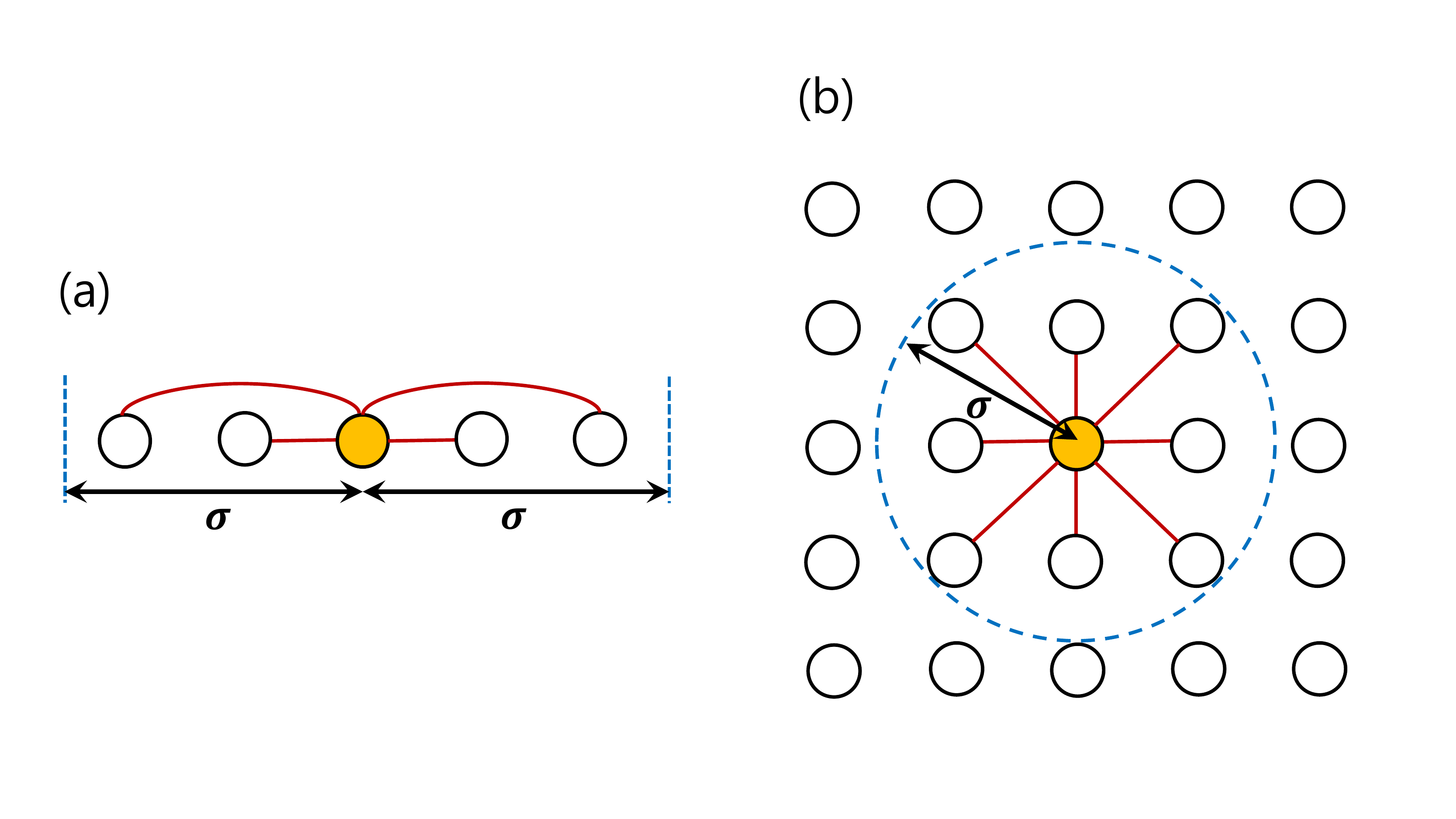}
\caption{(Color online) Schematic diagram for the network construction with interaction range $\sigma$.
Each node in (a) one- and (b) two-dimensional Euclidean lattices is connected 
to the nodes within the distance $\sigma$. Links connected to a representative node (colored yellow) are presented.     
}\label{figure1}
\end{figure}

In this section, we describe the system that we study in this paper. 
At first, we introduce how we construct the network structure. 
To reflect the finite distance between interacting individuals in real systems, 
we consider geometric distance in our model. We use $d$-dimensional Euclidean lattices with linear length $L$ for $d=1,2$ and
$N=L^d$ nodes are laid on the lattice sites. 
We define the distance between two nodes $i$ and $j$ by $|i-j|$ for $d=1$,
and by $\sqrt{(x_i-x_j)^2+(y_i-y_j)^2}$ for $d=2$, where $(x_i, y_i)$ are coordinates of node $i$. 
Then, we add a link between a pair of nodes if the distance between the pair of nodes is less than or equal to external parameter $\sigma$.
As $\sigma$ increases, each node is connected to the farther nodes and 
the network becomes all-to-all coupled structure (mean-field limit) as $\sigma \rightarrow L$.
Schematic diagram for this network construction is depicted in Fig.~\ref{figure1}.

Next, we describe the dynamic rule of the basic model~\cite{Hong} in this network structure.
In the basic model, two opposite opinions $A$, $B$ and neutral position $AB$ exist. 
Then, each node is in one state among the four discrete states $A$, $A_c$, $B$, and $AB$. Here, $B$ and $AB$ are states of having opinions on $B$ and $AB$.
Both $A$ and $A_c$ are states having opinion on $A$. However, each node in the state $A$ can change to $AB$
when it interacts with other nodes in the state $B$ while each node in the state $A_c$ never changes its state.
In other words, each node in the state $A_c$ maintains its state throughout the whole process. 
In this respect, each node in the state $A_c$ is called {\it{zealot}} (of opinion on $A$).

At the beginning of the dynamics, the fraction $p$ of nodes are randomly selected to be in state $A_c$ and 
all the other nodes are in state $B$. At each time step, we select a link randomly and choose 
one individual randomly from two ends of the selected link to be {\it{speaker}} and the other {\it{listener}}.
Then, the state of listener at the next time step is determined according to the rules described in Table.~\ref{table1}.
In this manner, the states of the nodes are updated and the system reaches equilibrium.
We use $n_A$, $n_B$, and $n_{AB}$ to denote the fractions of population for subpopulations in states $A$, $B$, and $AB$ at arbitrary time, respectively.
Therefore $n_A + n_B + n_{AB} + p = 1$ throughout the whole process.

In the previous study of the basic model in the mean-field limit~\cite{Hong}, it was reported that
the system arrives at the consensus on $A$ (i.e. $n_A = 1-p$ and $n_B=n_{AB}=0$) when $p>p^{MF}_c \equiv 1-\sqrt{3}/{2}\approx 0.134$.
On the contrary, when $p < p_c^{MF}$, majority of the population remains in the state $B$. Interestingly, the transition to consensus on $A$ is found to be discontinuous at $p_c^{MF}$~\cite{Hong}. This result implies that the zealots with the opposite opinion overturn the majority opinion in real systems. We now extend the previous study with the mean-field model, by controlling the interaction range $\sigma$ to reflect the finite-range interaction of real systems. 

\begin{table}
\includegraphics[width=1.0\linewidth]{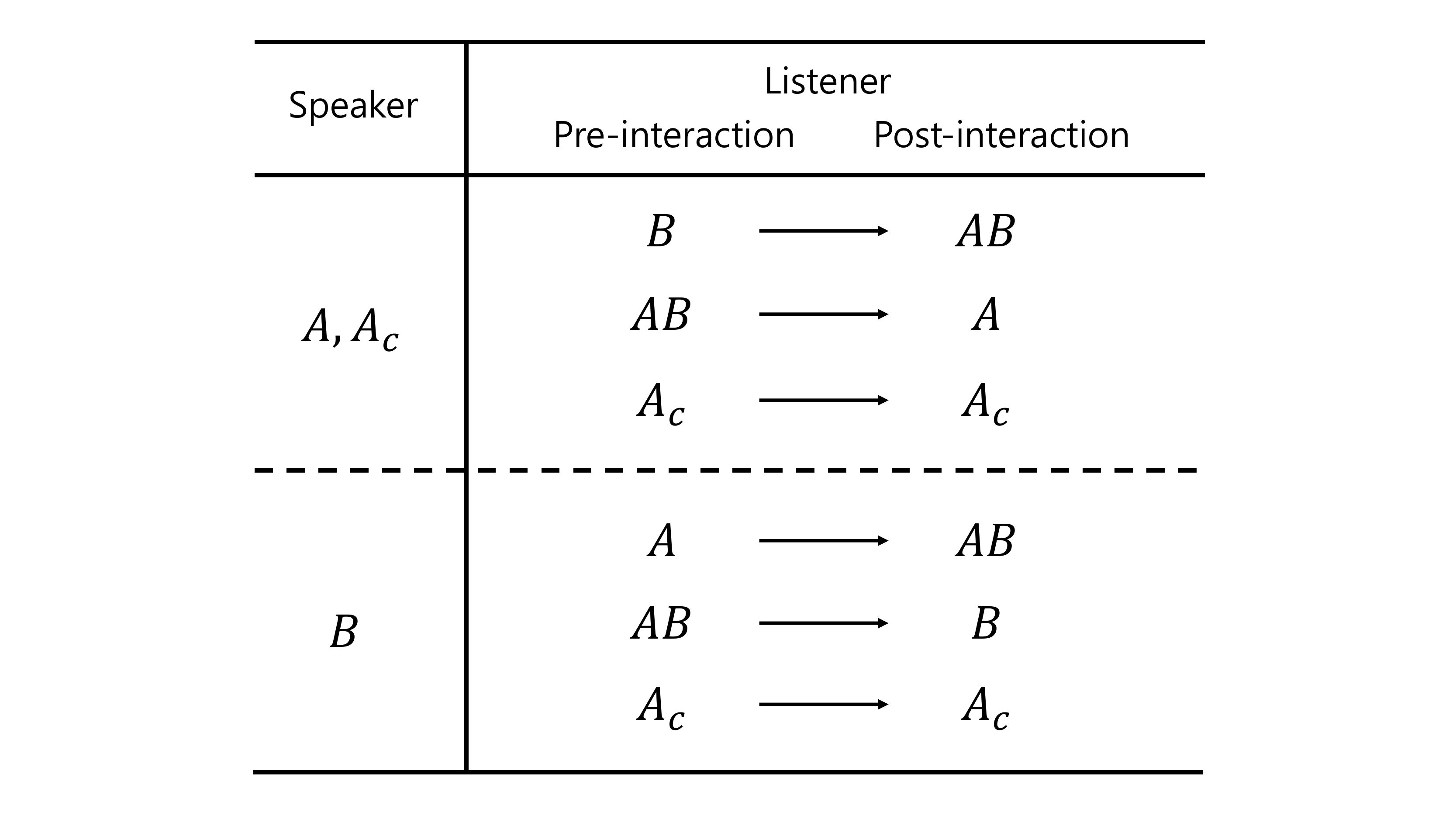}
\caption{List of changes of state of listener after interaction with speaker~\cite{Hong}. 
We note that zealots do not change their states irrespective of state of the speaker.  
}\label{table1}
\end{table}

\section{Numerical Analysis}

\begin{figure}
\includegraphics[width=1.0\linewidth]{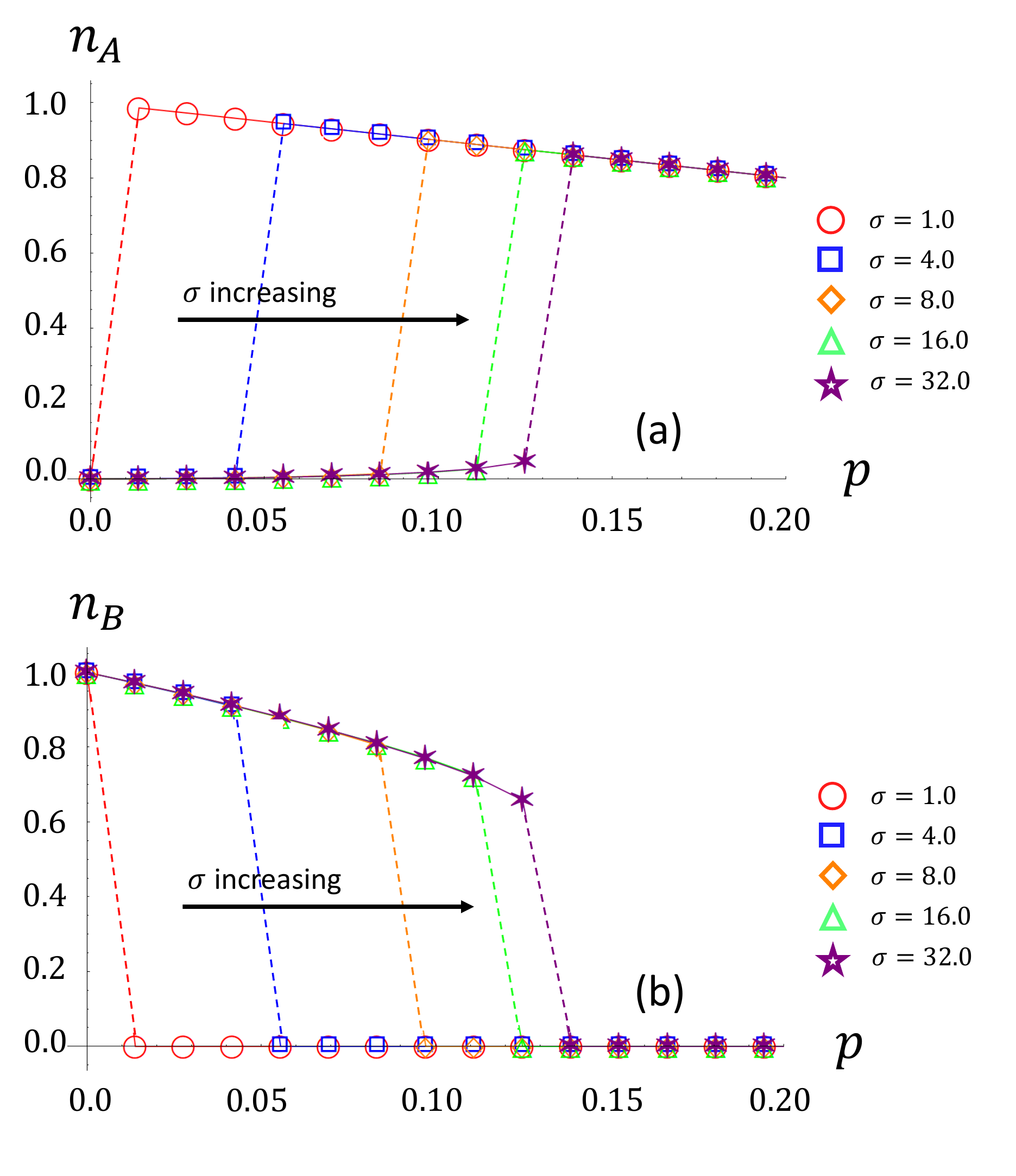}
\caption{(Color online) (a) $n_A$ and (b) $n_B$ vs $p$ in two dimensional lattice with $L=60$. We find that $p_c$ at which $n_A$ jumps to $1-p$ and $n_B$ drops to $0$ increases with $\sigma$.
The lines are guides to the eyes. 
}\label{figure2}
\end{figure}

To investigate the effects of the interaction range $\sigma$ on the opinion consensus dynamics, 
we numerically explore the system, performing the extensive simulations for various values of $\sigma \in [1, L)$.
At first, we measure the equilibrium values of $n_A$ and $n_B$ as functions of $p$ by increasing $\sigma$ from $\sigma=1$
to inspect how they approach to the results in the mean-field limit as $\sigma \rightarrow L$. 
In Fig.~\ref{figure2}, the result for $d=2$ is shown. 
Irrespective of $\sigma$, we find that discontinuous transition to consensus on $A$ occurs at $p=p_c$.
Moreover $p_c$ increases as $\sigma$ increases, as a result,
$p_c \rightarrow p^{MF}_c$ for $\sigma \rightarrow L$.
We observe the same tendencies of $n_A$ and $n_B$ for $d=1$.

\begin{figure}
\includegraphics[width=1.0\linewidth]{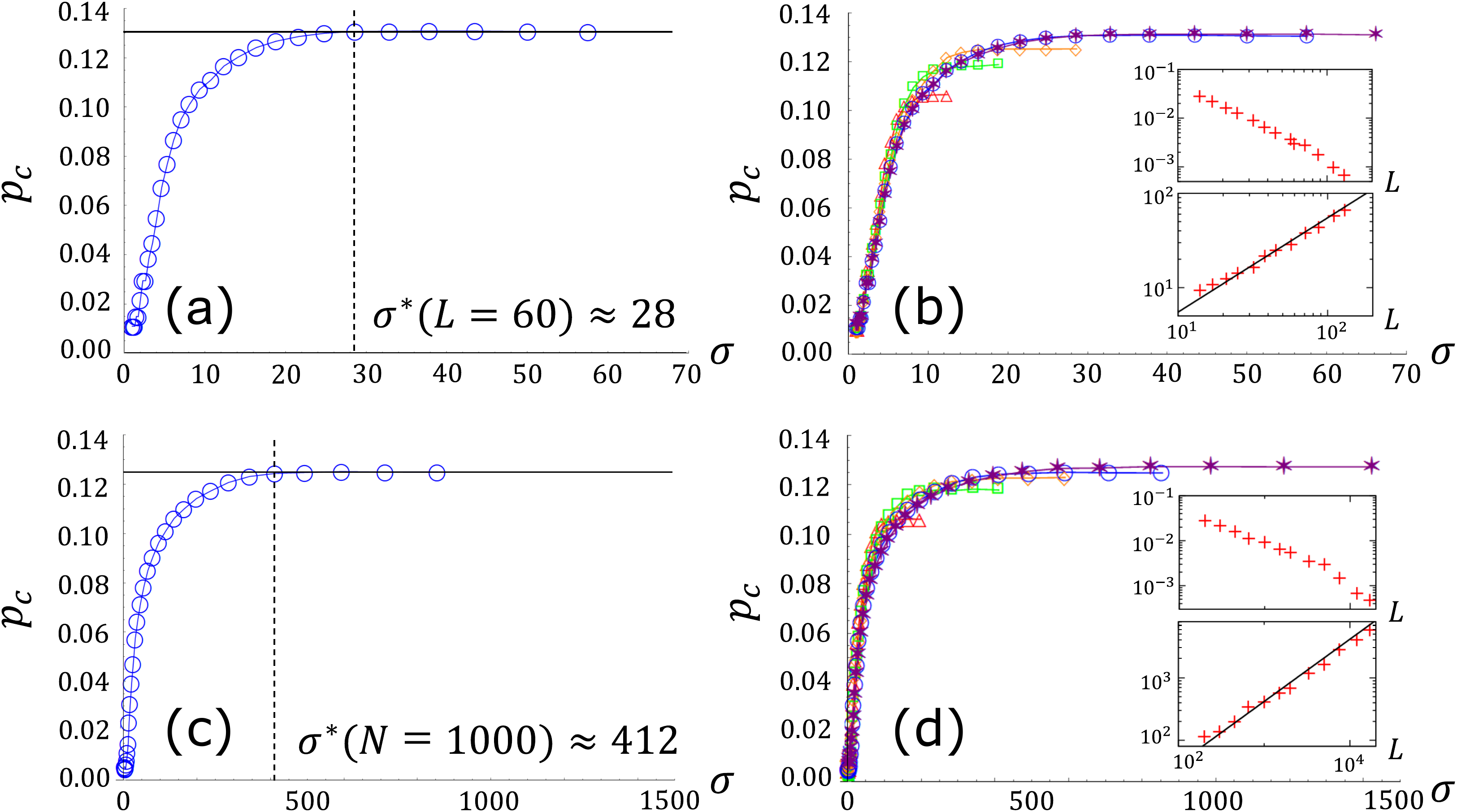}
\caption{(Color online) (a) $p_c$ vs $\sigma$ in two dimensions for $L=60$. 
We estimate $\sigma^* \approx 28$ in this system.
(b) $p_c$ vs $\sigma$ in two dimensions for various system sizes $L=14$, $21$, $32$, $60$, and $71$ from the left.    
Inset: (upper inset) $p_c^{MF} - p_c(\sigma^*)$ and (lower inset) $\sigma^*$ vs $L$.
The data of the upper inset indicates that $p_c(\sigma^*)$ increases to $p_c^{MF}$ as $L$ increases.
The slope of the guideline in the lower inset is $1$.
(c) $p_c$ vs $\sigma$ in one dimension for $N=1000$. 
We estimate $\sigma^* \approx 412$ in this system.
(d) $p_c$ vs $\sigma$ in one dimension for various system sizes $N/10=20$, $45$, $65$, $100$, and $150$ from the left.    
Inset: (upper inset) $p_c^{MF} - p_c(\sigma^*)$ and (lower inset) $\sigma^*$ vs $L$.
The data of the upper inset indicates that $p_c(\sigma^*)$ increases to $p_c^{MF}$ as $L$ increases.
The slope of the guideline in the lower inset is $1$.
}\label{figure3}
\end{figure}

Numerical results in Fig.~\ref{figure2} show that $p_c$ is strongly dependent on $\sigma$. 
To inspect the behavior of $p_c$ depending on $\sigma$ in more detail,
we plot numerically measured $p_c$ as a function of $\sigma$.
The results for $d=1$ and $d=2$ are shown in Fig.~\ref{figure3}.
For both spatial dimensions, the $p_c$ increases drastically as $\sigma$ increases 
in small $\sigma$ region while $p_c$ is constant in large $\sigma$ region.
Based on the shape of the curve $p_c$ vs $\sigma$, we estimate crossover point $\sigma^*$
of two different behaviors of $p_c$ for increasing $\sigma$, where
$p_c$ increases prominently for $\sigma < \sigma^*$ while it is constant
for $\sigma \geq \sigma^*$. 
We guess that $\sigma^*(N) \propto L$ based on numerical data in Fig.~\ref{figure3}.

The increasing behavior of the $p_c$ for $\sigma < \sigma^*$ seems to be 
counter-intuitive because the zealots with the opposite opinion would persuade majority of population more easily
as the interaction range increases. However, in our model, the zealots occupy smaller fraction than the $B$ at the beginning of the dynamics.
This initial condition is believed to make the influence of $B$ increase more rapidly than that of $A_c$ as the interaction range increases.
As the result, more zealots are required to reach the consensus on $A$ as the interaction range $\sigma$ increases.

\section{difference of dynamics to consensus between $\sigma<\sigma^*$ and $\sigma>\sigma^*$ regions}

\begin{figure}
\includegraphics[width=1.0\linewidth]{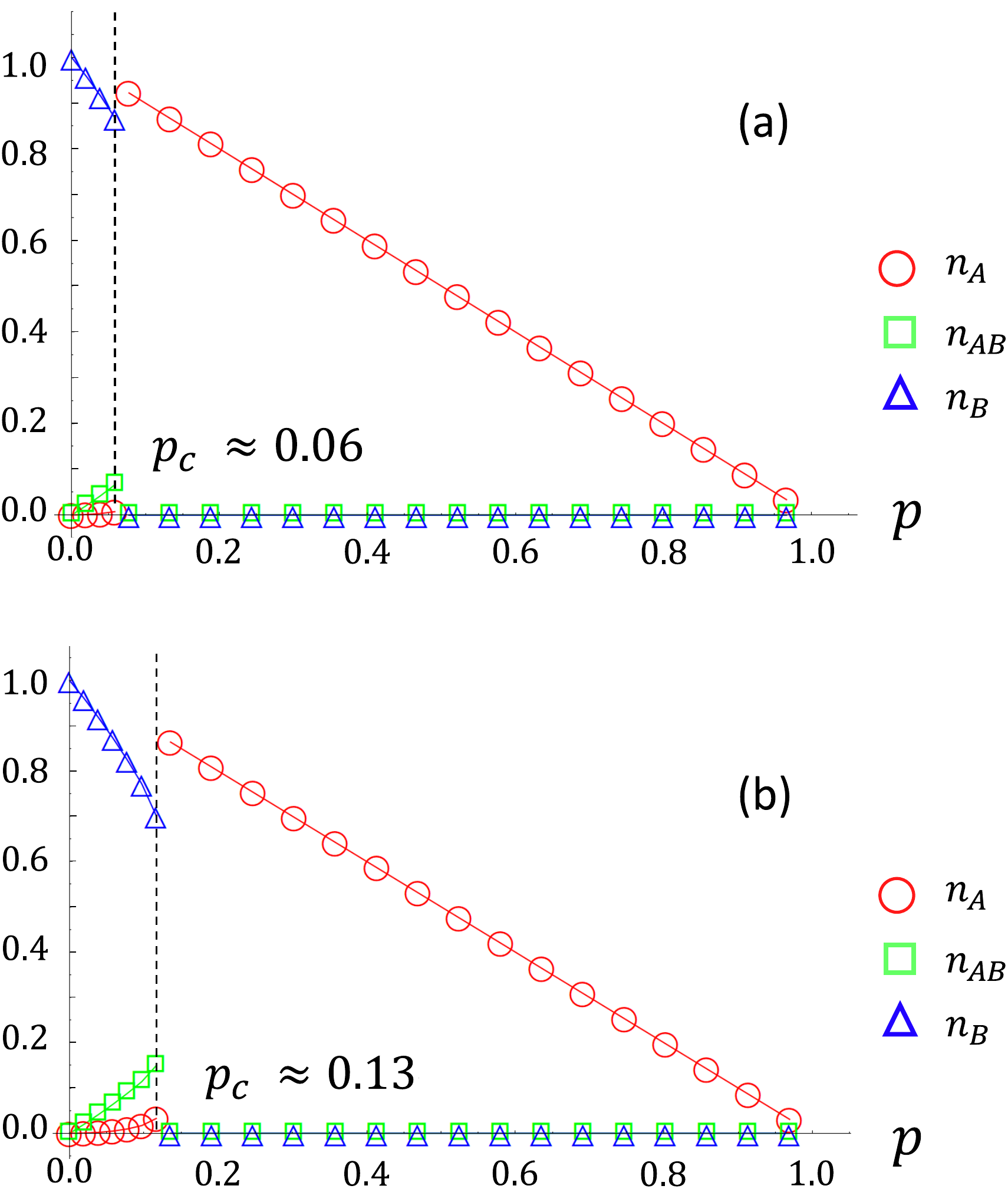}
\caption{(Color online) $n_A$, $n_{AB}$, and $n_B$ vs $p$ for (a) $\sigma=4 \ll \sigma^*$ and (b) $\sigma=32 > \sigma^*$ 
in two dimensions with $L=60$. $p_c \ll p_c^{MF}$ in (a) 
while $p_c \approx p_{c}^{MF}$ in (b). Black dashed line indicates the critical point $p_c$. 
}\label{figure4}
\end{figure}

In the previous section, 
we showed that $p_c$ increases as $\sigma$ increases 
for $\sigma <\sigma^*$ and $p_c$ is constant for $\sigma \geq \sigma^*$.
Fig.~\ref{figure4} shows the behavior of $n_A$, $n_B$, and $n_{AB}$ 
as a function of $p$. We find that the behaviors of the subpopulations for $\sigma > \sigma^*$
are similar to those for mean-field system in~\cite{Hong}, while the behaviors for $\sigma \ll \sigma^*$ are different from those.

\begin{figure*}
\includegraphics[width=1.0\linewidth]{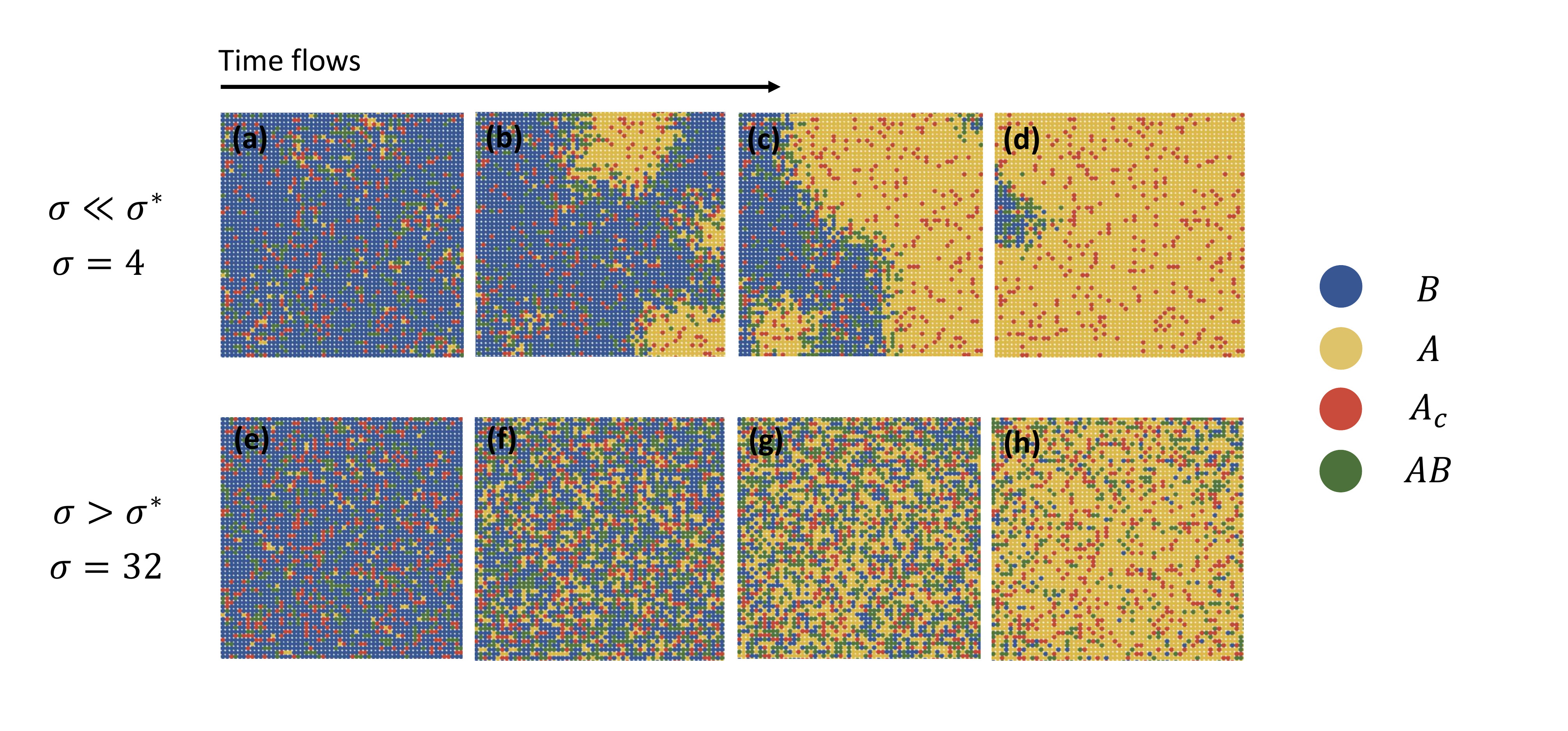}
\caption{(Color online) Snapshots during the evolution to the consensus on $A$ in two dimensions with $L=60$ for (a-d) $\sigma=4<\sigma^*$ and (e-h) $\sigma=32>\sigma^*$. 
To raise consensus on $A$ for both cases, we use $p=0.08 > p_c=0.06$ for (a-d) and $p=0.14 > p_c \approx p^{MF}_c$ for (e-h).
}\label{figure5}
\end{figure*}

To see the difference between two regions of $\sigma$ in more detail, we investigate the time evolution of population's states for $d=2$. 
The snapshots presenting the distributions of states at several instants of time for each of the two representative values of $\sigma$ are shown in Fig.~\ref{figure5}. The difference between the two cases is clearly displayed.

For $\sigma\ll \sigma^*$, each $A_c$ generates localized organization composed of $A$ states at early times.
As time goes on, each localized organization increases in size and 
two distinct organizations merge if they are adjacent to each other. In this manner, 
each organization composed of $A$ and $A_c$ states is compact.
For $\sigma > \sigma^*$, however, $A_c$ prefers to change distant neighbors into $A$ because 
the number of neighbors increases proportional to distance from the $A_c$. Therefore, 
states $A$ are scattered and their number increases as time goes on.

\section{Conclusion}

In social networks, each individual interacts with finite number of people contrary to the mean-field case. To reflect this fact, we introduce the network structure allowing finite-range interactions defined on one- and two-dimensional Euclidean lattices. We study the basic model on this network structure 
to describe the dynamics of opinion consensus in real systems.
As the interaction range increases, the critical fraction of zealots required to raise consensus on their opinion increases and approaches to its mean-field value. 
We present the different patterns of consensus formation between small and large interaction ranges. 
Especially, when the interaction range is small, zealots increase the community of their side by persuading the 
near ones, which would reflect the real systems.
It would be interesting to consider more realistic network structures where short-range interactions and a few long-range interactions coexist.

\begin{acknowledgments}
This paper was supported by research funds for newly appointed professors of Chonbuk National University in 2016 (Y. S. C),
the NRF Grants No. 2017R1C1B1004292 (Y.S.C) and No. 2018R1A2B6001790 (H.H).
\end{acknowledgments}

\end{document}